\documentclass[acmsmall]{acmart}

\setcopyright{cc}
\setcctype{by}
\acmDOI{10.1145/3715786}
\acmYear{2025}
\acmJournal{PACMSE}
\acmVolume{2}
\acmNumber{FSE}
\acmArticle{FSE067}
\acmMonth{7}
\received{2024-09-13}
\received[accepted]{2025-01-14}

\usepackage{textcomp}
\usepackage{csquotes}
\usepackage{amsfonts}
\usepackage{xspace}
\usepackage[ruled,linesnumbered,noend,]{algorithm2e} 

\SetKwInOut{Input}{input}
\SetKwInOut{Output}{output}
\SetKwProg{Fun}{function}{}{end}
\SetFuncSty{algoFuncSty}
\SetNlSty{textmd}{}{}

\SetCommentSty{commentSty}
\usepackage{tabularx}
\usepackage{booktabs}
\newcolumntype{L}{>{$}l<{$}}
\newcolumntype{R}{>{$}r<{$}}
\newcolumntype{C}{>{$}c<{$}}
\newcolumntype{D}{>{$\displaystyle}l<{$}}
\usepackage{amsmath}
\usepackage{hyperref}
\usepackage[capitalize]{cleveref}
\usepackage{multirow}
\usepackage{listings}
\usepackage{scratch3}
\setscratch{scale=.75}
\usepackage{subcaption}
\usepackage{tikz}
\usepackage[
]{siunitx}
\usepackage[inline]{enumitem}
\usepackage{framed}
\usepackage{adjustbox}

\usepackage{todonotes}
\makeatletter
\newcommand{\done}[1][]{\@ifnextchar[{\@donewitharg}{\@donewithoutarg}}
\newcommand{\@donewitharg}[2][]{\ignorespaces}
\newcommand{\@donewithoutarg}[1]{\ignorespaces}
\makeatother

\usepackage{float}

\newcommand{\toolname}[1]{{#1}}
\newcommand{\whisker}{\toolname{Whisker}\xspace}
\newcommand{\Scratch}{\toolname{Scratch}\xspace}

\newcommand{\evosuite}{\toolname{EvoSuite}\xspace}

\DeclareMathOperator{\ad}{ad}
\DeclareMathOperator{\aad}{aad}
\DeclareMathOperator{\failed}{failed}
\DeclareMathOperator{\passed}{passed}
\DeclareMathOperator{\totalFailed}{total failed}

\DeclareMathOperator{\total}{total}

\DeclareMathOperator*{\argmax}{arg\,max}

\newcommand{\hatblock}{%
  \begin{scratch}[print,fill blocks,fill gray=0.85]
    \blockinit{}
\end{scratch}}

\newcommand{\stackblock}{%
  \begin{scratch}[print,fill blocks,fill gray=0.85]
    \blockmove{}
\end{scratch}}

\newcommand{\cblock}{%
  \begin{scratch}[print,fill blocks,fill gray=0.85]
    \blockif{}{\blockspace[.4]}
\end{scratch}}

\newcommand{\capblock}{%
  \begin{scratch}[print,fill blocks,fill gray=0.85]
    \blockstop{}
\end{scratch}}

\newcommand{\reporter}{%
  \setscratch{print,fill blocks,fill gray=0.85}
  \ovalsound{\makebox[3.5em]{\raisebox{0pt}[1ex]{}}}
\setscratch{print,fill blocks,fill gray=0.85=false}}

\newcommand{\boolreporter}{%
  \setscratch{print,fill blocks,fill gray=0.85}
  \booloperator{\makebox[3em]{}}
\setscratch{print,fill blocks,fill gray=0.85=false}}

\newcommand{\dataset}[1]{\textsc{#1}\xspace}
\newcommand{\simple}{\dataset{Simple}}
\newcommand{\complex}{\dataset{Complex}}

\newcommand{\chkCovLvl}[1]{\textsf{#1}\xspace}
\newcommand{\blockSus}{\chkCovLvl{blk}}
\newcommand{\scriptSus}{\chkCovLvl{scr}}
\newcommand{\targetSus}{\chkCovLvl{act}}

\newcommand{\cumuChk}{\chkCovLvl{cumu}}
\newcommand{\testChk}{\chkCovLvl{test}}
\newcommand{\assChk}{\chkCovLvl{asrt}}

\newcommand{\fsInit}{\chkCovLvl{init}}
\newcommand{\fsSol}{\chkCovLvl{sol}}
\newcommand{\fsAll}{\chkCovLvl{all}}

\newcommand{\rqone}{How well can individual bugs be localized?\xspace}
\newcommand{\RQONE}{How Well Can Individual Bugs Be Localized?\xspace}
\newcommand{\rqtwo}{How well can individual bugs be repaired?\xspace}
\newcommand{\RQTWO}{How Well Can Individual Bugs Be Repaired?\xspace}
\newcommand{\rqthree}{How well can student solutions be repaired?\xspace}
\newcommand{\RQTHREE}{How Well Can Student Solutions Be Repaired?\xspace}

\newcommand{\summary}[2]{%
  \vspace{-0.3cm}%
  \begin{center}%
    \colorbox{gray!20}{%
      \parbox{.985\linewidth}{
        \textbf{\textsf{Summary (\textit{#1})}:}~%
        #2%
      }%
    }%
  \end{center}%
}

\newcolumntype{T}{>{\ttfamily\small}l}

\newcommand{\nodejs}{Node.js v18.18.0\xspace}
\newcommand{\debian}{Debian~11\xspace}

\newcommand{\nSimpleProjs}{\num{12}\xspace}
\newcommand{\nComplexProjs}{\num{3}\xspace}
\newcommand{\nComplexVariants}{\num{82}\xspace}

\newcommand{\nTestsSimple}{\num{54}\xspace}
\newcommand{\nAssertsPerTestSimple}{%
    \num[round-mode=figures,round-precision=3]{2.111111111}\xspace}
\newcommand{\nTestsWithMoreThanOneAssertSimple}{\num{17}\xspace}

\newcommand{\nAssertBoatRace}{\num{57}\xspace}
\newcommand{\nTestsBoatRace}{\num{32}\xspace}
\newcommand{\nAssertPerTestBoatRace}{%
    \num[round-mode=figures,round-precision=3]{1.78125}\xspace}
\newcommand{\nTestsWithMoreThanOneAssertBoatRace}{\num{19}\xspace}

\newcommand{\nAssertFruitCatching}{\num{160}\xspace}
\newcommand{\nTestsFruitCatching}{\num{28}\xspace}
\newcommand{\nAssertPerTestFruitCatching}{%
    \num[round-mode=figures,round-precision=3]{5.714285714}\xspace}

\newcommand{\nAssertSpaceship}{\num{38}\xspace}
\newcommand{\nTestsSpaceship}{\num{20}\xspace}
\newcommand{\nAssertPerTestSpaceship}{%
    \num[round-mode=figures,round-precision=3]{1.9}\xspace}
\newcommand{\nTestsWithMoreThanOneAssertSpaceship}{\num{11}\xspace}

\newcommand{\popSize}{\num{48}\xspace}
\newcommand{\nParallelEvals}{\num{12}\xspace}
\newcommand{\timeLimitEval}{\SI{60}{\second}\xspace}
\newcommand{\timeLimitGA}{\SI{200}{\minute}\xspace}
\newcommand{\timeLimitOnePlusOne}{\SI{40}{\hour}\xspace}

\newcommand{\nChecking}{\num{3}\xspace}
\newcommand{\nSuspect}{\num{3}\xspace}
\newcommand{\nTechniquesFL}{\num{81}\xspace}
\newcommand{\nReps}{\num{15}\xspace}
\newcommand{\flAlpha}{\SI{5}{\percent}}

\newcommand{\runtimeFixedMax}{\SI[round-mode=figures,round-precision=3]{15.237}{\second}\xspace}
\newcommand{\runtimeBuggyMax}{\SI[round-mode=figures,round-precision=3]{32.5605}{\second}\xspace}

\newcommand{\runtimeBuggyAvg}{\SI[round-mode=figures,round-precision=3]{8.089695454545454}{\second}\xspace}

\newcommand{\nMetrics}{9\xspace}
\newcommand{\bestMedExamScore}{\num[round-mode=figures,round-precision=3]{0.2636363636363636}\xspace}
\newcommand{\worstMedExamScore}{\num[round-mode=figures,round-precision=3]{0.47560975609756095}\xspace}
\newcommand{\bestMedExamScorePerProj}{\num[round-mode=figures,round-precision=3]{0.1388888888888889}\xspace}
\newcommand{\bestMedExamScoreProj}{StutteringMovement\xspace}
\newcommand{\worstMedExamScorePerProj}{\num[round-mode=figures,round-precision=3]{0.9558823529411765}\xspace}
\newcommand{\worstMedExamScoreProj}{MessageNeverReceived2\xspace}
\newcommand{\tournamentBestRank}{72\xspace}
\newcommand{\tournamentDuels}{3240\xspace}

\newcommand{\bestFullFixRateSimple}{\num[round-mode=figures,round-precision=3]{0.938889}\xspace}
\newcommand{\worstFullFixRateSimple}{\num[round-mode=figures,round-precision=3]{0.633333}\xspace}
\newcommand{\bestPartFixRateSimple}{\num[round-mode=figures,round-precision=3]{0.988889}\xspace}
\newcommand{\worstPartFixRateSimple}{\num[round-mode=figures,round-precision=3]{0.861111}\xspace}

\begin{document}

\title{RePurr: Automated Repair of Block-Based Learners’ Programs}

\author{Sebastian Schweikl}
\orcid{0000-0001-8037-2653}
\affiliation{%
  \institution{University of Passau}
  \city{Passau}
  \country{Germany}
}
\email{sebastian.schweikl@uni-passau.de}

\author{Gordon Fraser}
\orcid{0000-0002-4364-6595}
\affiliation{%
  \institution{University of Passau}
  \city{Passau}
  \country{Germany}
}
\email{gordon.fraser@uni-passau.de}

\begin{abstract}
  Programming is increasingly taught using dedicated block-based
  programming environments such as Scratch.  While the use of blocks
  instead of text prevents syntax errors, learners can still make
  semantic mistakes implying a need for feedback and help. Since
  teachers may be overwhelmed by help requests in a classroom, may not
  have the required programming education themselves, and may simply
  not be available in independent learning scenarios, automated hint
  generation is desirable. Automated program repair can provide the
  foundation for automated hints, but relies on multiple assumptions:
  (1) Program repair usually aims to produce localized patches for
  fixing single bugs, but learners may fundamentally misunderstand
  programming concepts and tasks or request help for substantially
  incomplete programs.
  (2) Software tests are required to guide the search and to localize
  broken statements, but test suites for block-based programs are
  different to those considered in past research on fault localization
  and repair: They consist of system tests, where very few tests are
  sufficient to fully cover the code. At the same time, these tests
  have vastly longer execution times caused by the use of animations
  and interactions on Scratch programs, thus inhibiting the
  applicability of metaheuristic search.
  (3) The plastic surgery hypothesis assumes that the code necessary
  for repairs already exists in the codebase. Block-based programs
  tend to be small and may lack this necessary redundancy.
  In order to study whether automated program repair of block-based
  programs is nevertheless feasible, in this paper we introduce, to
  the best of our knowledge, the first automated program repair
  approach for Scratch programs based on evolutionary search.
  Our RePurr prototype includes novel refinements of fault
  localization to improve the lack of guidance of the test suites,
  recovers the plastic surgery hypothesis by exploiting that a
  learning scenario provides model and student solutions as
  alternatives, and uses parallelization and accelerated executions to
  reduce the costs of fitness evaluations.
  Empirical evaluation of RePurr on a set of real learners' programs
  confirms the anticipated challenges, but also demonstrates that the
  repair can nonetheless effectively improve and fix learners'
  programs, thus enabling automated generation of hints and feedback
  for learners.
\end{abstract}



\maketitle

\section{Introduction}\label{sec:introduction}

Programming education is increasingly more wide spread and starts at
ever younger ages. To support learners with their initial endeavours
in programming, block-based programming environments such as
Scratch~\cite{resnick2009} simplify programming by replacing the
challenge of creating syntactically valid textual code with stacking
visual blocks of different shapes and colors. Even though resulting
programs are always syntactically valid, they may still be
semantically wrong, for example when learners misunderstand
programming concepts or the programming task they are supposed to
complete, or when they are simply stuck. In these cases educators need
to provide hints and feedback to support learners, but teachers may be
facing many help requests simultaneously in a classroom setting, may
not possess the necessary skills to effectively debug and understand
the learners' problems, and in independent learning scenarios there
simply are no educators that could be asked for help. Consequently,
there is a need for automated hint generation techniques.

Automated program repair (APR) is a promising approach to enable this
feedback. Given a broken learners' program, APR can generate a fix,
from which feedback can be derived in any number of ways ranging from
hinting at erroneous locations, or misunderstood concepts, to
providing the actual solution.
Determining the best way to present hints is an educational research
question, whereas this paper focuses on the technical side, providing
the foundation for such research.
While research on APR has made
significant progress in the context of regular, text-based programming
languages, this is not the case for block-based programming
languages---to the best of our knowledge, APR has not been
investigated for block-based programming at all yet.

Block-based programs are challenging for APR for multiple reasons:
\begin{enumerate}
  \item Prior research on APR commonly targets single faults requiring
    small, local changes, but learners may misunderstand programming
    concepts, thus producing fundamentally broken programs, and
    they may request help for substantially incomplete programs.
  \item Automated tests are required not only to validate generated
    patches, but also to localize faulty statements as candidates for
    modification, and to guide search algorithms in producing
    patches. Tests for block-based programs written in languages like
    \Scratch, however, are usually system
    tests~\cite{testing-scratch-automatically}, and since \Scratch
    programs tend to be small, very few tests are required to fully
    cover a program. In addition, test executions take very long, since
    \Scratch programs tend to contain long-running animations and visual
    effects, which is inhibitive for test-based APR.
  \item The plastic surgery hypothesis~\cite{barr2014} assumes the
    code needed for repairs already exists in the codebase and to
    create a fix the relevant fix ingredients just need to be copied to
    the right place. \Scratch programs tend to be small and typically
    lack this redundancy.
\end{enumerate}

In this paper we introduce the first APR framework for the popular
block-based programming environment Scratch, following the
generate-and-validate paradigm to search for patches that fix broken
programs such that all tests pass. To address the challenges caused by
block-based programs, our RePurr prototype implementation integrates
several optimizations:
Rather than producing many variants with small changes like in
template-based or cloze-style learning-based repair approaches, our
framework relies on metaheuristic search in order to evolve fixes even
for fundamentally broken or incomplete programs.
We refine spectrum-based fault localization to consider information at
the assertion-level rather than the test level, and devise a refined
fitness function that provides more fine-grained guidance.
To recover the plastic surgery hypothesis, we exploit other potential
fix sources that exist in programming education scenarios, such as
model solutions or candidate solutions from other learners.
Finally, RePurr accelerates and parallelizes test executions in order
to reduce the costs of fitness evaluations.

In summary, the contributions of this paper are as follows:
\begin{itemize}
  \item We introduce RePurr, the first automatic repair prototype for
    block-based languages, targeting \Scratch programs.
  \item We conduct the first empirical study of applying fault
    localization to block-based programs, and refine fault localization
    by deriving computation rules at the level of test assertions that
    better guide the search and generalize beyond \Scratch.
  \item We define mutation and crossover operators for repairing
    \Scratch programs, and introduce the idea of fix sources as a
    generalization of the plastic surgery hypothesis.
  \item We introduce a refined fitness function that considers how many
    assertions of a test were executed, as well as the difference
    between expected and actual values of failed assertions.
  \item We empirically evaluate the repair framework using 12
    single-fault programs and 82 real faulty programs created by
    children, studying how well fault localization works
    on block-based programs, whether the plastic surgery hypothesis can
    be approximated with alternative fix sources, and whether the repair
    scales to real broken programs created by learners.
\end{itemize}

Our experiments confirm the anticipated challenges, but demonstrate
that the search-based repair can improve and fix learners' programs,
despite the degraded performance of fault localization on block-based
programs, and even when applied to fundamentally broken learners'
programs.


\section{Background}\label{sec:background}

\subsection{Anatomy of a \Scratch Program}\label{subsec:scratch}

\Scratch is a block-based programming language where programs are assembled visually rather than textually. \Scratch
programs are developed and executed in the \Scratch IDE,\footnote{\url{https://scratch.mit.edu/projects/editor/}} which runs
in the web browser. In analogy to a theater, every \Scratch program consists of the \emph{stage}, which is displayed in
the IDE as a background image, and any number of actors (\emph{sprites}), which are shown as smaller pictures layered on
top of the stage.
Every sprite can be given its own behavior and state, which makes it similar to a class in an object-oriented programming language. The stage
implements global behavior and state. The behavior is structured in terms of scripts, which are pieces of
encapsulated \Scratch code, similar to methods in an object-oriented programming language. Sprites and the stage can contain any number of
scripts. Every script is made of \Scratch blocks, which serve as basic building blocks. \Scratch blocks are the
counterpart of statements and expressions in a textual language.
The IDE provides a drawer of predefined \Scratch blocks,
which
can
be dragged around and snapped together. Every block has a certain color and shape. The color is a visual aid that
indicates the broader meaning of the block (e.g., all blocks related to \enquote{motion} are blue), while the
\emph{shape} dictates which blocks can be connected, similar to puzzle pieces. This prevents syntax errors and makes
\Scratch more beginner-friendly.

\begin{table}
  \centering
  \scriptsize
  \caption{Summary of the block shapes present in \Scratch}
  \label{tab:block-shapes}

  \begin{tabular}{lll}
    \toprule
    Shape         & Name             & Purpose                                              \\ \midrule
    \hatblock     & Hat block        & Event handling, triggering script execution          \\ \addlinespace
    \stackblock   & Stack block      & General purpose, stateful computations, side effects \\ \addlinespace
    \cblock       & C block          & Structuring control flow: loops and conditions       \\ \addlinespace
    \capblock     & Cap block        & Stopping a script or the program                     \\ \addlinespace
    \reporter     & Regular reporter & String or number expressions                        \\ \addlinespace
    \boolreporter & Boolean reporter & Boolean expressions                                 \\ \addlinespace \bottomrule
  \end{tabular}
\end{table}

\Cref{tab:block-shapes} gives an overview of all block shapes in \Scratch.
The blocks shaped like a hat, stack, C, and cap
are statements.
They have notches and matching cut-outs, such that they can be composed sequentially.
They will be executed in linear order.
Hat blocks can only be placed at the beginning (top) of scripts, while cap blocks can only appear at the end (bottom).
A script must start with a hat block, otherwise it is \enquote{unconnected} (i.e., dead code), but it need not end in a cap block.
C blocks are compound statements that wrap other statements in their \enquote{mouth}.
Reporter blocks (expressions) are ovals or hexagons,
and cannot be composed sequentially.
Statements and expressions may have oval or hexagonal holes to take expressions of the matching shape as input.

\subsection{The \whisker Testing Framework for \Scratch}

\whisker~\cite{testing-scratch-automatically} is a framework for testing \Scratch
programs automatically using system tests consisting of sequences
of user actions.
Like the \Scratch IDE, \whisker is implemented using web technologies. It can be run in a browser
or in headless mode as a command line application.
A \whisker test suite~$T$ is a JavaScript program that contains at least one \whisker test~$t$. Every such test is a
JavaScript function that controls the execution of the \Scratch program~$p$ under test by sending user inputs to $p$ and
triggering events. A test has access to an API that allows users to send such inputs, inspect the internal
state of $p$, and make assertions.
An example is shown
in Listing~\ref{lst:whisker-test}.

\begin{figure*}
  \centering

  \captionof{lstlisting}{JavaScript code of an example \whisker test~\lstinline|t| that checks a \Scratch
  program~\lstinline|p|}

  \label{lst:whisker-test}

    \begin{lstlisting}[gobble=4,basicstyle=\ttfamily\scriptsize]
    async function t(p) {
        const bowl = p.getSprite("Bowl");                      // Retrieve the "Bowl" sprite
        const bowlX = bowl.x;                                  // Inspect its internal state
        p.inputImmediate({device: "keyboard", key: "Right"});  // Send input events
        await p.runForTime(200);                               // Control program execution
        p.assert.ok(bowl.x > bowlX);                           // Assert that the bowl moved
    }
    \end{lstlisting}

  \vspace{-1em}
\end{figure*}


\section{Automatic Repair of \Scratch Programs}\label{sec:scratch-automatic-repair}

Given that we potentially need to fix incomplete and arbitrarily
broken programs, we use a repair approach based on genetic
programming, in which programs are evolved towards functional
correctness as determined by a set~$T$ of test cases. At the core of the
repair approach is a meta-heuristic \emph{search algorithm}
(\cref{subsec:algorithm}) which controls the exploration of the space
of possible programs, represented as their abstract syntax trees.
This is feasible because the target audience of \Scratch are
beginners, who tend to write comparatively small programs.  The
meta-heuristic search is based on evolutionary algorithms, where
typically a population of candidate individuals is gradually modified
and recombined. A \emph{fitness function}
(\cref{subsec:fitness-function}) determines which individuals are
selected for reproduction, and these are subjected to \emph{crossover}
(\cref{subsec:crossover}) and \emph{mutation} (\cref{subsec:mutation}).
These modifications are informed about how likely individual aspects
of the program are related to the test failures using \emph{fault
localization} techniques (\cref{subsec:fault-localziation}), and by
\emph{fix sources} (\cref{subsec:fix-sources}) from which to sample
possible modifications. We implemented this approach in the RePurr
prototype, which is an extension of the \whisker testing framework for
\Scratch.

\subsection{Repair Algorithm Outline}\label{subsec:algorithm}


{
  \scriptsize
  \begin{algorithm}
    \DontPrintSemicolon

    \SetKw{Break}{break}

    \SetKwFunction{viable}{filter\_viable}
    \SetKwFunction{sel}{select\_parents}
    \SetKwFunction{xov}{cross\_over}
    \SetKwFunction{mut}{mutate}
    \SetKwFunction{fit}{evaluate\_fitness}
    \SetKwFunction{best}{update\_elite}

    \Input{program~$p$ to repair, test suite~$T$, fix source~$F$}
    \Input{maximizing fitness function~$f_T$, population size~$n$, elitism size~$m$}
    \Output{repaired program~$p'$}

    \BlankLine

    $P \gets E \gets \{ p \}$\; \label{line:init}

    \BlankLine

    \While{stopping condition not reached}{ \label{line:evo-loop}

        $O \gets \emptyset$\;

        \BlankLine

        \While{$|O| + |E| < n$}{
            \uIf{$P = \{ p \}$}{
                $O \gets O \cup \{ \mut_F(p) \}$\; \label{line:init-pop}
            }
            \Else{
                $(p_1, p_2) \gets \sel_{f_T}(P)$\; \label{line:sel}
                $(o_1, o_2) \gets \xov(p_1, p_2)$\; \label{line:xov}

                \BlankLine

                \ForEach{$o \in \{ o_1, o_2, p_1, p_2 \}$}{
                    $O \gets O \cup \{ \mut_F(o) \}$\; \label{line:mut}

                    \BlankLine

                    \If{$|O| + |E| = n$}{ \label{line:break}
                        \Break\;
                    }
                }
            }
        }

        \BlankLine

        $P \gets \viable_{f_T}(O \cup E)$\; \label{line:next-pop}

        $E \gets \best_{f_T}(P, m)$\; \label{line:update-elite}
    }

    \BlankLine

    \Return{$p' \in \argmax_{p' \in E} f_T(p')$} \label{line:return}


  \caption{Repair algorithm for \Scratch}
    \label{alg:repair}
  \end{algorithm}

}

\noindent%
While there are many different techniques for APR~\cite{gazzola2019}
in general, these have not been explored on Scratch. We therefore
establish a baseline using the approach of GenProg~\cite{genprog},
which is among the most well known techniques, and has popularized the
use of evolutionary methods for APR. Inspired by this,
\cref{alg:repair} shows how our elitist genetic algorithm drives the
search for correct program versions, with adaptations to suit the
\Scratch domain.
The algorithm takes as input the program~$p$ to repair, a fix source~$F$ that governs where mutations may draw fixes
from (see \cref{subsec:fix-sources}), and a \whisker test suite~$T$ that encodes the expected behavior of the program.
Besides these inputs, the algorithm also takes a maximizing fitness function~$f_T$ based on the test suite~$T$, a
population size~$n \geq 2$, and an elitism size~$e < n$.

The algorithm evolves a population~$P$ of chromosomes, which represent potential repairs of the original program~$p$,
but we also maintain a subset $E \subset P$ of the most promising repairs found so far (the \emph{elite}). At the
beginning in \cref{line:init}, both $P$ and $E$ are formed by only $p$ itself.
In the evolutionary loop starting at \cref{line:evo-loop}, an offspring population~$O$ is derived from $P$. This is done
in two ways.

The first time only, an \emph{initial population} is created in \cref{line:init-pop}. It contains random mutants of $p$,
generated by applying the mutation operator $\mut_F$ to it, until $O$ contains $n-1$ mutants. The algorithm then skips ahead
to \cref{line:next-pop}, where an unmodified copy of $p$ (the current elite) is added to the population. This is to
ensure that the genetic material of $p$ is not lost and that no solution worse than $p$ will be returned. While $\mut_F$
is designed to produce syntactically valid mutants, one has to expect different program semantics and non-functional
properties, both in positive and negative ways, compared to $p$. For example, bugs might have been fixed, but a mutant
might also show excessive memory consumption or runtime, or other unrecoverable errors. Therefore, $\viable_{f_T}$
performs a viability check by running the test suite~$T$ against every program variant and evaluating its fitness via
the fitness function $f_T$. Inviable chromosomes for which fitness could not be determined (as explained in
\cref{subsec:fitness-function} and discussed in \cref{subsec:results-rq3}) are removed.
This forms the initial population~$P$. Finally, in \cref{line:update-elite} the function $\best_{f_T}$ takes the
fittest $m$ chromosomes of $P$ to become the new elite~$E$.

From now on, the evolutionary loop always creates offspring~$O$ from $P$ as follows. In \cref{line:sel}, parents are
chosen for reproduction by $\sel_{f_T}$ using rank-based selection on their fitness values. Then,
like in GenProg,
parents are probabilistically crossed over via \xov in \cref{line:xov} before being probabilistically mutated
alongside their offspring
by $\mut_F$ in \cref{line:mut}. This is repeated until there are $n - m$ offspring (\cref{line:break}). As before,
offspring~$O$ and elite~$E$ are combined and their viability ensured (\cref{line:next-pop}) to form the \emph{next
generation}~$P$. Because $\viable_{f_T}$ can remove chromosomes, $P$ might contain fewer than $n$ entries. Finally, the
elite is updated again in \cref{line:update-elite} with the best chromosomes in $P$.
The evolutionary loop repeats itself until the stopping condition is reached, e.g., the program has been fixed or a
user-defined time limit is reached, before \cref{line:return} returns a program variant~$p'$ with the best fitness from
the current elite.


\subsection{Fitness Function}\label{subsec:fitness-function}

We treat program repair as maximization problem: A test suite~$T$ encodes the expected behavior of the repair
subject~$p$, and the aim of the search is to produce a program variant~$p'$ that maximizes the number of passing tests.
The global optimum is reached when all tests $t \in T$ pass, in which case the program is considered fully repaired.
Note that even a partial repair can be sufficient in order to generate meaningful hints. For instance, users may only
request help for
\begin{enumerate*}
\item one specific failing test, or
\item a specific location where they are stuck.
\end{enumerate*}

\subsubsection{Infrastructure}

Before the fitness of $p$ can be computed, we must run $T$ against it. While the search operators have been
carefully designed to produce syntactically and grammatically valid \Scratch programs, the semantics could be
arbitrarily broken. For example, $p$ could run indefinitely (e.g., due to infinite loops) or consume an inordinate
amount of memory (e.g., due to recursive cloning.) To handle this, we evaluate $p$ in its own context,
which isolates it from evaluations of other programs. The steps are as follows:
\begin{enumerate*}
\item Create a new evaluation context,
\item load the \whisker testing framework,
\item load the project~$p$ and test suite~$T$ into \whisker, and
\item run $T$ against $p$.
\end{enumerate*}

In general, program repair incurs high computational effort, which is exacerbated by the time-based nature of \Scratch
and the long run times of \whisker tests~\cite{scratch-automated-test-generation}. To handle this, we restructured
parts of \whisker to make parallel test executions possible. The heap size for each evaluation is limited to
\SI{5}{\giga\byte}. When a fitness evaluation exceeds a given time limit, we terminate it by killing the associated
process, and reclaim the consumed resources.
Because recursive cloning can trigger a bug%
\footnote{\url{https://github.com/scratchfoundation/scratch-vm/issues/714}}
in the Scratch VM that immediately freezes it, we also implemented a
mechanism that detects and kills such zombie processes.
Implicitly, all operations in \cref{alg:repair} with a subscript $f_T$ use this infrastructure to evaluate fitness. As
this may be done several times per generation for the same chromosome, we cache fitness evaluations to avoid costly
redundant evaluations. Type-2 code clone detection could be used to save further evaluations in the future.

The result of a test execution is a trace that contains
\begin{enumerate*}
\item the test status,
\item the set of passed assertions, and
\item the assertion distance (\cref{subsubsec:assertion-distance}) of the failed assertion, if any.
\end{enumerate*}
Traces further contain coverage information to enable fault localization (\cref{subsec:fault-localziation}).

\subsubsection{Execution Status}

A passing test indicates that a bug was fixed or no regression was
introduced. Every passing test increases the fitness by $1$. A failing test can be the result of a failed assertion
(e.g., caused by a regression in $p$) or another error (e.g., the test reached the evaluation timeout.)

\subsubsection{Passed Assertions}

When a test fails, it is usually because the assertion last executed failed. Since a test can contain multiple
assertions, two programs that fail the same test can still have different failed assertions. As this information
provides more fine-grained feedback and potentially better gradients, we count
the number of passed assertions per failed test, and divide it by the total number of assertions in the test. This
ratio is always smaller than $1$, and estimates \enquote{how much} of the test the program passed. For
example, the ratio will be higher for a program that fails only the last assertion, compared to a program that
fails the first assertion. Higher ratios result in better fitness. The ratio is $1$ if all assertions (i.e., the
entire test) pass.

\subsubsection{Assertion Distance}
\label{subsubsec:assertion-distance}

In the context of test generation, branch distance~\cite{branch-distance} is a~common metric applied to branch conditions in the
program under test. It estimates \enquote{how close} control flow comes to taking the opposite branch, and is used
to guide the search towards yet unexplored parts of the program. For example, the branch distance of
\lstinline|if (5 <= 4)| is $|5-4|=1$, but for \lstinline|if (10 <= 4)| it is higher ($|10 - 4| = 6$)
because the latter condition is \enquote{farther away} from \lstinline|true| than the former.

Every \whisker assertion, such as \lstinline|p.assert.ok(x <= y)|, can be rewritten as a branching condition, e.g.,
\lstinline|if (y < x) p.assert.fail()|. This makes it possible to compute the branch distance for assertions, which
we call \emph{assertion distance}. Unlike branch distance, the assertion distance is only computed for the failing
assertion of a test, not the branches in the program we want to repair. The intention is to guide the search towards
new program variants that pass an assertion that previously failed, which allows the next test assertion to be reached.

\whisker records the actual and expected values for a failed assertion, which enables the distance computation $\ad(a)$
for the assertions~$a$,
following the rules established in the
literature~\cite{branch-distance}. String comparisons are handled by computing their Levenshtein distance, and
distances for emptiness checks on array-like objects can be computed by comparing their \lstinline|length| property
against \lstinline|0|.
In general, operands are assumed to be of the same type, but when a string is compared against a number, we convert the
string into a number before computing the distance.
This decision was motivated by the fact that \Scratch often stores numeric content as strings, and is in line
with the implicit type conversions Scratchers are already familiar with.

Remaining cases not covered by these rules tend do be comparisons on
Boolean values (e.g., \lstinline|p.assert.isTrue(...)|). For Boolean
expressions such as $ (x < 42) \land (y \geq z) $ testability
transformations~\cite{harman2004,certainty-booleans} are required to
recover distance information. Boolean values are typically based on
expressions also evaluated in programs, such as sensing checks, e.g.,
to check if a sprite touches a color
(\setscratch{scale=0.6}\boolsensing{touching color \ovalnum{} ?}) or
another sprite (\boolsensing{touching \ovalsensing*{sprite} ?}).
Our current
prototype does not yet implement testability transformations, such
that in these cases the assertion distance defaults to $\infty$.

\subsubsection{Fitness Formula}

As explained before, the range of the assertion distance is $(0, \infty]$. Its value gets smaller the closer the
assertion condition is to \lstinline|true|. As we want to combine it with the number of passing tests and passed
assertions, we require two adjustments.
First, we use the established function~$\alpha$~\cite{branch-distance-arcuri} to normalize the distance, such that its range is now $(0, 1]$.
Second, a program that comes closer to passing an assertion should be given higher fitness. We achieve this by
subtracting the normalized distance from $1$. This leads to the definition of the \emph{adjusted} assertion
distance~$\aad$:
\begin{equation*}
  \aad(a) = 1 - \alpha(\ad(a))
  \quad
  \text{where}
  \quad
  \alpha(x) =
  \begin{cases}
    1,           & \text{if $x = \infty$,} \\
    x / (x + 1), & \text{otherwise.}
  \end{cases}
\end{equation*}
For an execution $t(p)$ of the test $t$ against a \Scratch program $p$ the fitness~$f'_t(p)$ is given as
\begin{equation*}
  f'_t(p) =
  \begin{cases}
    1,                                                & \text{if $t(p)$ passed,} \\
    \frac{\passed(t(p)) + \aad(a)}{\total(t)}, & \text{if an assertion $a$ of $t(p)$ failed.}
  \end{cases}
\end{equation*}
Here, \enquote{$ \passed $} and \enquote{$ \total $} refer to the number of passed and total assertions, respectively.
The value of $ \aad(a) $ is smaller than $1$, such that a passing assertion contributes more to the fitness than the
assertion distance. In turn, the fraction $ (\passed + \aad) / \total $ is smaller than $1$, such that it is outweighed
by a passing test. This is to avoid deceiving fitness landscapes.

Finally, the fitness~$f_T(p)$ of~$p$ given the execution~$T(p)$ of a test suite~$T$ is computed as the sum of the
fitnesses of its test executions:
\(
  f_T(p) = \sum_{t \in T} f'_t(p)
\).

\subsection{Crossover}\label{subsec:crossover}

The crossover operator exchanges sprites, including all their scripts, between two \Scratch programs. It is applied with
a probability of \SI{70}{\percent}. 
When applied, it takes two parent programs $p_1$ and $p_2$, and creates two offspring $o_1$ and $o_2$.
In detail, the operator creates a bijective mapping $ M \subset S_1 \times S_2 $ between
the sprites $S_1$ of $p_1$ and $S_2$ of $p_2$, such that code is exchanged for functionally similar sprites. This is
done by computing the similarity between all pairs of sprites, and picking the pairs with the highest similarity to
become the matching pairs $M$. If there are two candidate pairs $(s, x)$ and $(s, y)$ of highest similarity and $x \neq
y$ we pick one randomly. Note that $|S_1| = |S_2|$ because our search operators never create new sprites or delete
existing sprites. This ensures every sprite is matched exactly once. Because certain blocks cannot be used on the stage,
such as \enquote{move}
blocks, our algorithm always matches the stage of $p_1$ with the stage of $p_2$ to prevent malformed programs. Finally,
a randomly chosen subset $M' \subset M$ of matching sprites is exchanged between the parents $p_1$ and $p_2$, resulting
in offspring $o_1$ and $o_2$. We require $M' \neq \emptyset$ and $M' \neq M$ as this would create offspring identical to
their parents.
We measure
the similarity of sprites using the Levenshtein distance between their names. This is done because sometimes students
might misspell words.
An alternative approach would be to match sprites based on the similarity of the syntax
trees. For example, the pq-gram distance~\cite{augsten2005} is often used in data-driven hint
generation~\cite{zimmerman2015} and has been proposed for matching sprites in \Scratch~\cite{obermueller2021}.


\subsection{Mutation}\label{subsec:mutation}

In contrast to crossover, the mutation operator works on the block level. Its definition is inspired by
\evosuite\cite{evosuite}. There are three operations, applied in this order:
\begin{enumerate*}
\item delete,
\item change,
\item insert.
\end{enumerate*}
Each is applied independently with probability $1/3$, such that on average only one of them is applied. With a small
probability of $ (1/3)^3 = \SI{3.7}{\percent} $ all are applied, while in $ (2/3)^3 = \SI{29.6}{\percent} $ of cases
none are applied. This means at least one operation is applied with probability \SI{70.4}{\percent}.

\subsubsection{Block Sampling}\label{subsubsec:sampling}

\newcommand{\prank}{\ensuremath{p}}

Every operation uses fault localization (\cref{subsec:fault-localziation}) to preferably act on suspicious blocks.
It uses this information to create a ranked list of blocks, from rank~$n$ (most suspicious) to rank~$1$ (least
suspicious).
Blocks of equal suspiciousness share the same rank. 
Then, a rank~$r$ is selected with probability~$\prank(r)$ directly proportional to~$r$:
\(
  \label{eq:prank}
  \prank(r) 
  = \frac{r}{\sum_{i=1}^{n}i}
  = \frac{2 \times r}{n \times (n + 1)}
  = r \times \prank(1)
\).
Lastly, a block of rank~$r$ is chosen with uniform probability.
This process can be repeated to sample multiple blocks. When suspiciousness is given on the script
or sprite level, the process is analogous, and once a script or sprite has been chosen, we select one of its blocks with
uniform probability.

\subsubsection{Delete}

\begin{figure}
  \centering
  \begin{minipage}[b]{.49\textwidth}
    \centering
    \begin{subfigure}[b]{.49\columnwidth}
      \centering
      \begin{scratch}[scale=0.5]
        \blockrepeat{repeat until
          \setscratch{print,fill blocks,fill gray=1}
          \boolsensing{touching \ovalsensing*{edge} ?}
          \setscratch{print=false}
        }{
          \blockmove{move
            \setscratch{print,fill blocks,fill gray=1}
            \ovaloperator{\ovalnum{5} + \ovalvariable{my variable}}
            \setscratch{print=false}
          steps}
        }
      \end{scratch}
      \caption{Before deletion}
    \end{subfigure}
    \hfill
    \begin{subfigure}[b]{.49\columnwidth}
      \centering
      \begin{scratch}[scale=0.5]
        \blockrepeat{repeat until \boolempty[3em]}{
          \blockmove{move \ovalnum{} steps}
        }
      \end{scratch}
      \caption{After deletion}
    \end{subfigure}
    \caption{Deleting expressions creates unoccupied holes}
    \label{fig:delete-expr}
  \end{minipage}
  \hfill
  \begin{minipage}[b]{.49\textwidth}
    \centering
    \begin{subfigure}[b]{.49\columnwidth}
      \centering
      \begin{scratch}[scale=0.5]
        \blockinit{when \greenflag\ clicked}
        \setscratch{print,fill blocks,fill gray=1}
        \blockrepeat{repeat until \boolsensing{touching \ovalsensing*{edge} ?}}{
          \blockmove{move \ovaloperator{\ovalnum{5} + \ovalvariable{my variable}} steps}
        }
        \setscratch{print=false}
        \blockstop{stop \selectmenu{all}}
      \end{scratch}
      \caption{Before deletion}
    \end{subfigure}
    \hfill
    \begin{subfigure}[b]{.49\columnwidth}
      \centering
      \begin{scratch}[scale=0.5]
        \blockinit{when \greenflag\ clicked}
        \blockstop{stop \selectmenu{all}}
      \end{scratch}
      \caption{After deletion}
    \end{subfigure}
    \caption{Deleting a C block and its nested statements}
    \label{fig:delete-cblock-all}
  \end{minipage}
\end{figure}

\begin{figure}
  \centering
  \begin{minipage}[b]{.49\textwidth}
    \centering
    \begin{subfigure}[b]{.49\columnwidth}
      \centering
      \begin{scratch}[scale=0.5]
        \blockinit{when \greenflag\ clicked}
        \setscratch{print,fill blocks,fill gray=1}
        \blockrepeat{repeat until \boolsensing{touching \ovalsensing*{edge} ?}}{
          \setscratch{print=false}
          \blockmove{move \ovaloperator{\ovalnum{5} + \ovalvariable{my variable}} steps}
          \setscratch{print,fill blocks,fill gray=1}
        }
        \setscratch{print=false}
        \blockstop{stop \selectmenu{all}}
      \end{scratch}
      \caption{Before deletion}
    \end{subfigure}
    \hfill
    \begin{subfigure}[b]{.49\columnwidth}
      \centering
      \begin{scratch}[scale=0.5]
        \blockinit{when \greenflag\ clicked}
        \blockmove{move \ovaloperator{\ovalnum{5} + \ovalvariable{my variable}} steps}
        \blockstop{stop \selectmenu{all}}
      \end{scratch}
      \caption{After deletion}
    \end{subfigure}
    \caption{Deleting a C block without its nested blocks}
    \label{fig:delete-cblock-only}
  \end{minipage}
  \hfill
  \begin{minipage}[b]{.49\textwidth}
    \centering
    \begin{subfigure}[b]{.49\columnwidth}
      \centering
      \begin{scratch}[scale=0.5]
        \setscratch{print,fill blocks,fill gray=0.85}
        \blockinit{when \greenflag\ clicked}
        \setscratch{print=false}
        \setscratch{print,fill blocks,fill gray=1}
        \blockmove{New location 1}
        \setscratch{print=false}
        \setscratch{print,fill blocks,fill gray=0.85}
        \blockrepeat{repeat until \boolsensing{touching \ovalsensing*{edge} ?}}{
          \setscratch{print=false}
          \blockmove{move \ovaloperator{\ovalnum{5} + \ovalvariable{my variable}} steps}
          \setscratch{print,fill blocks,fill gray=0.85}
        }
        \setscratch{print,fill blocks,fill gray=1}
        \blockmove{New location 2}
        \setscratch{print=false}
        \setscratch{print,fill blocks,fill gray=0.85}
        \blockstop{stop \selectmenu{all}}
        \setscratch{print=false}
      \end{scratch}
      \caption{New location choices}
    \end{subfigure}
    \hfill
    \begin{subfigure}[b]{.49\columnwidth}
      \centering
      \begin{scratch}[scale=0.5]
        \blockinit{when \greenflag\ clicked}
        \blockmove{move \ovaloperator{\ovalnum{5} + \ovalvariable{my variable}} steps}
        \blockrepeat{repeat until \boolsensing{touching \ovalsensing*{edge} ?}}{
          \blockspace[.5]
        }
        \blockstop{stop \selectmenu{all}}
      \end{scratch}
      \caption{Moving to location~1}
    \end{subfigure}
    \caption{Moving the blue stack block to a new location}
    \label{fig:move-stmt}
  \end{minipage}
\end{figure}

Blocks $b_1, \ldots, b_l$ of the current project are sampled without replacement. Then, the first block $b_1$ is deleted
with probability $1/l$. The next block $b_2$ is deleted with probability $1/l$, and so on. If a
deleted block $b_i$ contains a set of nested expressions $\{ b_j \}_{j>i}$, they are deleted as well, potentially
leaving the block surrounding $b_i$ with an unoccupied hole, see \cref{fig:delete-expr}. \Scratch treats hexagonal holes
as the Boolean value \enquote{true}, and oval holes as the number~$0$ or the empty string, depending on the context.
Nested statements $\{ b_j \}_{j>i}$ of a deleted C block $b_i$ are handled in two ways, chosen with the same
probability: Either, they are deleted along with $b_i$ as shown in \cref{fig:delete-cblock-all}, or only $b_i$ is
deleted, which \enquote{unwraps} its nested statements as shown in \cref{fig:delete-cblock-only}.

\subsubsection{Change}

Blocks $b_1, \ldots, b_l$ are sampled without replacement. Then, the first block $b_1$ is changed with
probability $1/l$. The next block $b_2$ is changed with probability $1/l$, etc.

When a suspicious block~$b_i$ is changed, exactly one of four operations is performed, each with the same probability:
\begin{enumerate*}[label=\alph*)]
\item 
\emph{swapping} positions with another, randomly chosen block;
\item 
\emph{moving} it to a different, randomly chosen location;
\item 
\emph{replacing} it with the copy of another, randomly chosen block;
\item 
\emph{selecting} a different value in its dropdown menu, if present.
\end{enumerate*}
If $b_i$ is a compound block, the operation also applies to all its nested blocks.
For example, moving an if-then-else block also moves the condition and the two branches along with it.
In general, all operations can act on blocks located
anywhere in the program, unrestricted by script or sprite boundaries. In case of swap and replace, only blocks of
compatible shapes are used.

\Cref{fig:move-stmt} shows the effect of moving a nested statement block to a new location, leaving the mouth of the C
block empty. Usually, a statement can sit between any two existing statements (unless it is a hat or cap block).
In contrast, an expression can only be moved if there is a block with an unoccupied hole matching its shape.
Replacement combines the effects of deleting a block, copying a different block, and moving it to the unoccupied
location.
Moving an expression leaves a free hole behind where
other expressions can be inserted.

\subsubsection{Insert}

A statement $s_1$ is sampled from the set of statements $S$ of the repair subject with replacement, and a
random statement $r_1$ is copied (including its nested expressions) from the fix source~$F$ (\cref{subsec:fix-sources}),
before inserting $r_1$ at location $s_1$. Then, additional randomly chosen statements from $F$ are inserted with
logarithmically decreasing probability: the first additional statement~$r_2$ at a location $s_2$ sampled from $S \cup \{
r_1 \}$ with probability $\sigma$, the next statement $r_3$ at $s_3$ sampled from $S \cup \{r_1, r_2 \}$ with
probability $\sigma^{2}$, and so on, until the $i$th statement is not inserted.

Preferably, a new statement $r_1$ is inserted after the sampled statement $s_1$ or alternatively, in case of C blocks, as the first
statement of its mouth. Insertion before a statement is only considered for the first statement of an unconnected
script.
Insertions of C blocks are treated in a special way. With equal
probability, they are inserted either leaving their mouth empty, or
such that they wrap a stack of connected suspicious blocks. The stack
size is chosen randomly, and may range from a single statement to the
entire script.
As  expressions require an empty hole, they are not handled by insertion, but rather via replacement of empty holes.


\subsection{Fault Localization}\label{subsec:fault-localziation}

Execution traces contain the set of \Scratch blocks covered by the test, which can be used for spectrum-based fault
localization (SFL)~\cite{fault-loc-survey} to identify the blocks likely responsible for a bug. Our instrumentation
measures the coverage of statement blocks, and we consider an expression block covered if its parent statement is
covered (over-approximation).

\subsubsection{Metrics}

Various SFL metrics have been proposed, with Tarantula, Ochiai and DStar among the most
popular ones.
Each calculates a suspiciousness of a block~$b$ based on how often it was executed by failed vs.\ passed tests.
For example, the Ochiai metric calculates suspiciousness as:
\(
  \failed(b) / \sqrt{\totalFailed \times (\failed(b) + \passed(b))}.
\)
A higher number of failed and lower number of passed tests leads to higher
suspiciousness.  In total, we implemented 9 metrics:
Tarantula~\cite{tarantula}, Ochiai~\cite{ochiai}, Jaccard, Zoltar,
Op2~\cite{NaishLR11}, Kulczynski2~\cite{lourencco2004binary},
McCon~\cite{mcconnaughey1964determination}, Barinel~\cite{AbreuZG09},
and DStar2~\cite{dstar}.
Note that every SFL metric requires coverage information about $b$ to compute its suspiciousness. This information is
not available if
\begin{enumerate*}
\item the test suite did not cover $b$, or if
\item after the last test execution $b$ was inserted as a new block or at a different location by mutation or
  crossover.
\end{enumerate*}
In these cases, we use $-\infty$ as suspiciousness.

SFL tends to work best with a large number of unit tests, each of
which covers disjoint parts of the code, which help to isolate the
fault from correct program parts. \Scratch projects, however, are
comparatively small such that the number of tests is usually
limited. Furthermore, \whisker tests are essentially system tests,
which means that individual tests do not exercise isolated aspects of
the code, but tend to cover most parts of the program. Moreover,
projects often have multiple bugs and might even be missing parts of
the implementation.
We tried to address these challenges by extending the SFL definitions by suspect levels and checking levels.

\subsubsection{Suspect Levels}

Suspiciousness metrics can be calculated for different levels of
coarseness. We compute suspiciousness on the block level (\blockSus),
matching the common approach for statement-level fault localization.
Since SFL has not been explored previously on \Scratch programs and
can be expected to perform sub-optimally for the reasons outlined
above, we added two more suspect levels, \enquote{script} (\scriptSus)
and \enquote{sprite} (\targetSus), in order to determine the best
level of granularity empirically.
On the script level, the suspiciousness of
entire scripts instead of individual blocks is computed. That is, we only allow scripts as input $b$, and change
$\passed(b)$ and $\failed(b)$ to refer to the number of passed and failed tests that executed script $b$. The sprite
level is defined analogously.

\subsubsection{Checking Levels}

\begin{table}[t]
  \begin{minipage}[t]{.3\linewidth}
    \captionof{lstlisting}{Example test case $t$}
    \label{lst:chk-test-suite}
    \vspace*{-.5em}
        \begin{lstlisting}[mathescape,basicstyle=\ttfamily\scriptsize,gobble=12]
            async function $t$(p) {
                // Executed suspects: $B_1$
                p.assert.$a_1$;
                // Executed suspects: $B_2$
                p.assert.$a_2$;
                // Executed suspects: $B_3$
                p.assert.$a_3$;
            }
        \end{lstlisting}
  \end{minipage}
  \hspace*{.1\linewidth}
  \begin{minipage}[t]{.36\linewidth}
    \scriptsize
    \centering
    \caption{Coverage by checking level}
    \label{tab:checking-levels}
    \begin{tabular}{R*{7}{C}}
      \toprule
      & \multicolumn{3}{c}{\assChk} & \multicolumn{3}{c}{\cumuChk} & \testChk \\
      \cmidrule(lr){2-4} \cmidrule(lr){5-7} \cmidrule(lr){8-8}
      & a_1 & a_2 & a_3 & a_1 & a_2 & a_3 & t \\
      \midrule
      B_1 & \checkmark & & & \checkmark & \checkmark & \checkmark & \checkmark \\
      B_2 & & \checkmark & & & \checkmark & \checkmark & \checkmark \\
      B_3 & & & \checkmark & & & \checkmark & \checkmark \\
      \bottomrule
    \end{tabular}
  \end{minipage}
\end{table}

By default, suspiciousness metrics are calculated using
the number of passed and failed tests that covered a suspect~$b$. We
call this the \enquote{test} checking level (\testChk). To accommodate
for smaller test suites where individual tests tend to cover large
parts of the program, we added the more fine-grained checking levels
\enquote{assertion} (\assChk), and \enquote{cumulative} (\cumuChk). On
the assertion level, the functions check how many assertions covered
$b$ (rather than how many tests). An assertion $a_i$ covers $b$ if $b$
was executed since the previous assertion $a_{i-1}$. The cumulative
level is an extension of the assertion level, where an assertion $a_i$
covers $b$ if $b$ was executed since the beginning of the current
test, but it need not be executed between $a_{i-1}$ and $a_i$.
To illustrate the difference between checking levels, consider a \whisker test
case $t$ with three assertions $a_1$, $a_2$, $a_3$, and sets of suspects $B_1, B_2,
B_3$ executed between assertions as shown in Listing~\ref{lst:chk-test-suite}. The
coverage matrix of each checking level is shown in \cref{tab:checking-levels}.


\subsection{Fix Sources}\label{subsec:fix-sources}

The plastic surgery hypothesis is central to search-based program repair. It assumes
the blocks needed for a bug fix are already contained in the program itself; they just
need to be copied to the ``correct'' position. The mutation operators ``replace'' and
``insert'' described in \cref{subsec:mutation} were designed with this in mind.
However, this might not apply to small and incomplete implementations often produced by
young learners. For this reason, we tried to recover the hypothesis by extending the
sources a fix can be drawn from: In a classroom setting, teachers can provide sample
solutions, and alternative implementation attempts may be given by other students.

Prior work has considered these additional sources for hint generation, for example by proposing edits that make the program more similar to the model solution~\cite{Fein2022,obermueller2021,Obermueller2023}. Programming tasks, however, can usually be solved in different ways. If the model solution uses a
different approach, then resulting hints would force students towards the model solution
approach, rather than fixing the students' chosen approaches. One way to address this would be by providing different model solutions. However,
the problem is exacerbated in Scratch, where code can be arbitrarily distributed across
concurrent scripts. Student solution attempts could also be considered, but
hints based on incorrect solutions may be misleading. In contrast, repair
can explore new coding pathways not present in other solutions, while still optimizing
for functional correctness thanks to the use of test suites.

We define three fix sources for ``replace'' and ``insert'' operations:
\begin{enumerate*}
\item Our baseline is \fsInit, the plastic surgery hypothesis. It takes blocks from the original repair subject only.
\item With \fsSol, fixes can additionally be drawn from a sample solution.
\item The \fsAll fix source extends sol by also including alternative (possibly buggy) implementations.
  In other words, $ \fsInit \subset \fsSol \subset \fsAll $.
\end{enumerate*}

If multiple fix sources are given, our prototype chooses one with uniform probability, before extracting a block from it with uniform probability (stratified sampling.)
An alternative approach might be to rank fix sources based on their similarity to the program under repair, as is done in data-driven hint generation \cite{zimmerman2015}.



\section{Evaluation}\label{sec:experiments}

To gain a better understanding of the problem of automatic program repair for \Scratch we aim to empirically answer the following research questions for \Scratch programs:
\begin{description}
  \item[RQ1] \rqone
  \item[RQ2] \rqtwo
  \item[RQ3] \rqthree
\end{description}

\subsection{Experimental Setup}

\subsubsection{Implementation}\label{subsec:implementation}
RePurr implements the repair approach described in
\cref{sec:scratch-automatic-repair} as an extension of
\whisker~\cite{testing-scratch-automatically}.
As baselines for comparison we implemented a random search and a (1+1)EA, informed by
studies on Genetic Improvement suggesting that local search via stochastic hillclimbing
similar to our (1+1)EA performs well \cite{BlotP21}. Both baseline algorithms use the
same mutation operator as the genetic algorithm, but the mutation
probability is set to \SI{100}{\percent} such that in every iteration
a new variant is
created.
Random Search does this by always mutating the original program, while
(1+1)EA always mutates (and unless the mutation caused a decrease of
fitness, updates) the best variant found so far.
The source code is publicly available: \url{https://github.com/se2p/whisker}.

\subsubsection{Execution Environment}\label{subsec:environment}
We created a Docker image of \whisker based on \debian and \nodejs.
All experiments were conducted on a computing cluster consisting of 17 machines, each equipped with a 24-core AMD EPYC 7443P CPU
at \SI{2.85}{\giga\hertz}, and \SI{256}{\giga\byte} of RAM.

\subsubsection{Datasets}

\begin{table}[tb]
  \scriptsize
  \centering
  \sisetup{round-mode=figures,round-precision=3}
  \caption{Characteristics of the faulty program variants in the \simple dataset}
  \begin{tabular}{l*{6}{r}SS}
\toprule
Variant & Blocks & Scripts & Targets & Pass & Fail & Assertions & {Coverage [\%]} &  {Duration [s]} \\
\midrule
   ComparingLiterals1 & 34 &  5 & 3 & 2 & 1 &  5 & 86.2069 & 1.515 \\
   ComparingLiterals2 & 66 &  8 & 7 & 6 & 3 & 14 & 89.7959 & 2.604 \\
   ForeverInsideLoop1 & 35 &  5 & 5 & 2 & 1 &  7 & 90.3226 & 0.767 \\
   ForeverInsideLoop2 & 33 &  3 & 3 & 2 & 2 & 20 & 51.5152 & 2.010 \\
MessageNeverReceived1 & 46 &  7 & 5 & 5 & 2 & 10 & 94.5946 & 4.521 \\
MessageNeverReceived2 & 80 & 10 & 4 & 3 & 1 & 11 & 91.6667 & 1.516 \\
     MessageNeverSent & 56 & 10 & 6 & 5 & 1 & 11 & 73.4694 & 2.221 \\
    MissingCloneInit1 & 51 &  6 & 3 & 1 & 2 & 14 & 86.4865 & 0.868 \\
    MissingCloneInit2 & 77 & 13 & 5 & 2 & 1 &  5 & 54.2373 & 0.731 \\
  MissingLoopSensing1 & 44 &  6 & 3 & 3 & 1 &  6 & 89.1892 & 0.853 \\
  MissingLoopSensing2 & 49 &  6 & 5 & 4 & 1 &  6 & 86.1111 & 1.036 \\
   StutteringMovement & 39 &  7 & 3 & 2 & 1 &  5 & 85.2941 & 0.732 \\
\bottomrule
\end{tabular}


  \label{tab:simple}
\end{table}%

\begin{figure*}[tb]
  \centering
  \includegraphics[width=\linewidth]{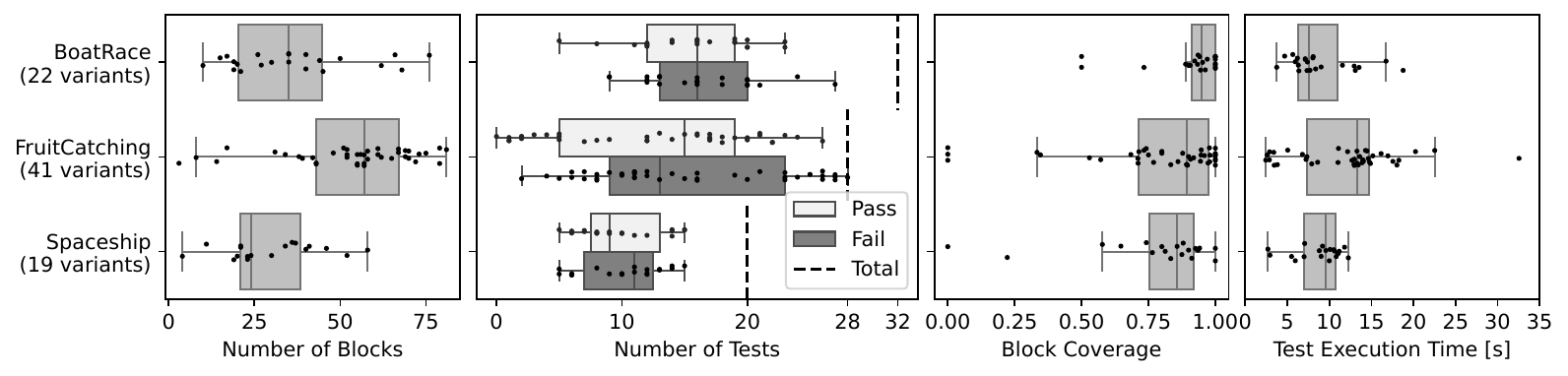}
  \caption{Characteristics of the faulty program variants in the \complex dataset}
  \label{fig:complex}
\end{figure*}%

Our study uses two datasets of \Scratch projects: 
The \simple dataset (\cref{tab:simple}) consists of \nSimpleProjs\ \Scratch
projects, each containing a different representative fault covering
different categories of common \Scratch
bugs~\cite{fraedrich2021}. Every fault in the dataset is fixable by
one or two mutations of a block, and the dataset also contains fixed
versions and \whisker test suites able to detect the bugs.
Overall, there are \nTestsSimple test cases, with \nAssertsPerTestSimple assertions on
average. Multiple assertions are present in \nTestsWithMoreThanOneAssertSimple tests,
and every test suite contains at least one such test case.
The \complex dataset (\cref{fig:complex}) consists of \nComplexProjs real-world
\Scratch projects from a classroom setting (BoatRace, FruitCatching,
and Spaceship.) For each of the projects there is one model solution
without any bugs, and multiple incomplete or faulty student solution
attempts.
These projects are bigger and more complex than the projects in the
\simple dataset, include faults of omission, and are generally more expensive to test.
Their test suites contain \nAssertBoatRace, \nAssertFruitCatching, and
\nAssertSpaceship assertions, respectively, resulting in \nAssertPerTestBoatRace,
\nAssertPerTestFruitCatching, and \nAssertPerTestSpaceship assertions per test on
average. For FruitCatching, all \nTestsFruitCatching tests contain multiple assertions,
while \nTestsWithMoreThanOneAssertBoatRace of \nTestsBoatRace do for BoatRace, and
\nTestsWithMoreThanOneAssertSpaceship of \nTestsSpaceship do for Spaceship.

\subsubsection{Hyperparameters}
Given the execution environment, our GA uses a population size of~\popSize, such that half a generation could be
evaluated on one machine in parallel. But given the less potent hardware in a classroom setting, we opted for
\nParallelEvals~parallel evaluations to make for a more realistic scenario. While \Scratch programs run at 30~FPS
natively, \whisker also offers accelerated execution. We used its maximum setting, rendering as many frames at
a time as possible, and measure the runtime in terms of wall-clock time.
With this, computing the fitness of a sample solution in the more expensive \complex dataset takes
less than \runtimeFixedMax on our hardware, while evaluating a buggy version takes
\runtimeBuggyAvg on average and up to \runtimeBuggyMax. Based on this, we specified a time limit of \timeLimitEval per
evaluation, and \timeLimitGA for the entire GA, such that it evolves at least 50 generations. RS uses the same
parameters. Because (1+1)EA cannot be parallelized, its time limit is \nParallelEvals~times as long, i.e.,
\timeLimitOnePlusOne.

\subsection{Methodology}\label{subsec:methodology}

\subsubsection{RQ1: \RQONE}

We manually determined the faulty blocks of the \nSimpleProjs~projects in the \simple dataset. Then, we computed the
suspiciousness of all blocks using the \nTechniquesFL~FL techniques obtained by combining the \nMetrics~SBFL metrics,
\nSuspect~suspect levels, and \nChecking~checking levels mentioned in \cref{subsec:fault-localziation}. Although
\whisker tries to ensure deterministic test executions, we repeated the experiment \nReps~times to accommodate for any
remaining flakiness.

The combinations are compared by their EXAM scores~\cite{exam-score}, which approximate
how many blocks of a program need to be inspected before the fault is localized.
Similar to the mutation operators in \cref{subsec:mutation}, all $N$ blocks of a
project are ranked by their suspiciousness values, using rank~$1$ for the most
suspicious, and rank~$N$ for the least suspicious block. The rank of the actually
faulty block is divided by $N$ to obtain the score, ranging from $1/N$ (best) to
$1$~(worst). If multiple blocks share the same suspiciousness, we use
their average rank.
For multi-block faults we take the faulty block ranked worst, assuming all faults must be localized before the defect
can be repaired. For faults of omitted statements we consider the preceding statement faulty. In case of omitted
expressions, we labeled the enclosing block faulty. This is motivated by the workings of the mutation operator, because it inserts statements \emph{after} and expressions \emph{into} a given insertion
point. When operating on the script or sprite level, each block is assigned the suspiciousness of its original script
or sprite, before computing the ranking.

Finally, the EXAM-scores of all combinations are compared pairwise in a tournament ranking. To this,
we check if the difference between two combinations is statistically significant using a Mann-Whitney $U$-test with
$ \alpha = \flAlpha $. If so, we then computed the Vargha Delaney $A_{12}$ effect size to find the better of the two
contestants. In the end, the combination that wins the most duels represents the best FL technique for \Scratch.

\subsubsection{RQ2: \RQTWO}

To answer this question, we apply the genetic algorithm (GA),
(1+1)EA, and random search (RS) as baseline to the 12 faulty programs from the
\simple dataset.
For fault localization, we always use the best technique according to
RQ1.  For each algorithm, we used the fix source \fsInit, followed by
a run with \fsSol.  Every experiment is repeated 15 times to
accommodate for randomness.
To answer the RQ we compare the fix rates. We distinguish between
partial fixes (at least one more test is passed than the initial
variant) and full fixes (all tests pass): While a full fix is of
course desirable, a partial fix is already sufficient in practice to
synthesize a hint for the learner on how to correct a mistake or how
to proceed with the programming task.

\subsubsection{RQ3: \RQTHREE}

The setup of RQ3 is similar to RQ2, but uses the \complex dataset of
\nComplexVariants real-world \Scratch programs.
As programs written by learners may not contain relevant code from
which to synthesize fixes, this RQ also investigates if additional fix
sources can recover the plastic surgery hypothesis. To this, we conduct an additional
run where mutations use the fix source \fsAll, besides \fsInit and \fsSol, respectively.
The experiment is repeated 15
times, for a total of $ 3 \times 3 \times 82 \times 15 = 11070 $ runs. To answer
the RQ, we consider the full and partial fix rates, the number of
additional passing tests per solution, and the time until the first
improvement was found.

\subsubsection{Threats to Validity}\label{subsec:threats-to-validity}

\emph{Threats to external validity} arise due to our sample of
\Scratch projects: The high computational costs and large number of
necessary comparisons and repetitions limit the feasible number of
faulty programs to be repaired, and although the sample of programs
originates from real educational contexts, they may not generalize.
\emph{Threats to construct validity} may result from our comparisons
based on partial and full fix rates, which may not be representative
for the effectiveness when synthesizing actual feedback to
learners. The EXAM score used for RQ1 may also not be representative
for incomplete and multi-fault solutions, but labelling actual fault
locations for programs in the \complex dataset would be tricky.
\emph{Threats to internal validity} may arise from errors in our
repair framework or experiment infrastructure. There may be other fix
sources (e.g., based on large language models), other mutation
operators, or further optimizations that would affect
results. However, the current repair framework represents best
practices in program repair and provides initial insights, on which
future research can improve.


\section{Results}\label{sec:results}

\subsection{\RQONE}\label{subsec:results-rq1}

\begin{table}
  \scriptsize
  \centering
  \caption{Tournament ranking of the \nTechniquesFL FL techniques by their EXAM scores}
\begin{tabular}{lllr}
\toprule
     Metrics & Susp. & Chk. & Wins \\
\midrule
     DStar2,
    Jaccard,
Kulczynski2,
      McCon,
     Ochiai,
         Op,
     Zoltar & \blockSus & \cumuChk &    72 \\
    Barinel,
  Tarantula & \blockSus & \cumuChk &    71 \\
         Op & \blockSus & \testChk &    63 \\
     DStar2,
    Jaccard,
Kulczynski2,
      McCon,
     Zoltar & \blockSus & \testChk &    62 \\
         Op,
    Barinel,
     Ochiai,
  Tarantula & \blockSus & \testChk &    56 \\
    Barinel,
     DStar2,
    Jaccard,
Kulczynski2,
      McCon,
     Ochiai,
  Tarantula,
     Zoltar & \blockSus & \assChk &    27 \\
Kulczynski2,
      McCon,
         Op,
     Zoltar,
         Op & \scriptSus & \testChk &    13 \\
     DStar2,
    Jaccard,
Kulczynski2,
      McCon,
     Ochiai,
         Op,
     Zoltar & \targetSus & \assChk &    13 \\
     DStar2,
    Jaccard,
Kulczynski2,
      McCon,
     Ochiai,
         Op,
     Zoltar & \scriptSus & \assChk &     2 \\
    \multicolumn{3}{l}{(All remaining techniques)} &     0 \\
\bottomrule
\end{tabular}

  \label{tab:exams-tournament}
\end{table}

\Cref{tab:exams-tournament} shows the results of ranking all \nTechniquesFL FL techniques by their EXAM scores in a
tournament. Of the \tournamentDuels duels, the vast majority ended in a draw, indicating no significant difference in
bug finding ability. The most wins (\tournamentBestRank) were achieved by \blockSus suspects and \cumuChk checking,
combined with DStar2, among other metrics which showed the same performance. Because DStar2 also performs well on real faults in other languages~\cite{eval-and-improve-fault-loc}, we focus the remaining discussion on it.

\begin{figure}
  \centering
  \begin{subfigure}[b]{.46\linewidth}
    \centering
    \includegraphics[width=\linewidth]{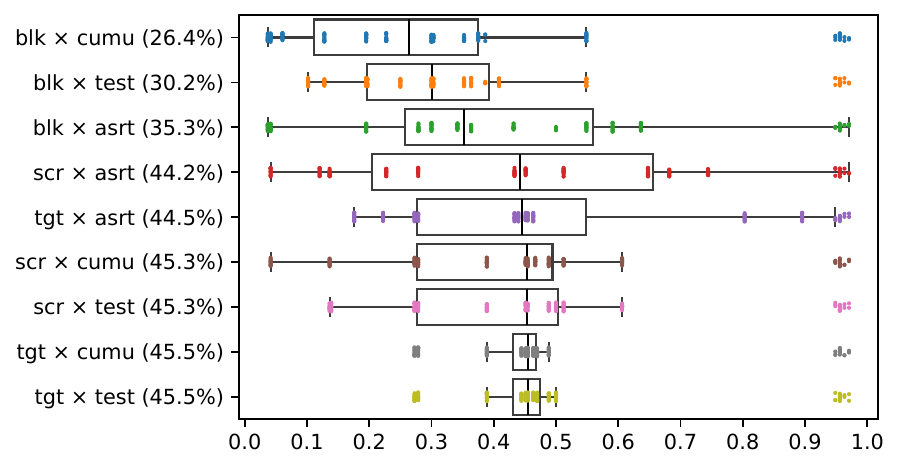}
    \caption{Per suspect level and checking level}
    \label{fig:exams-dstar2-per-combination}
  \end{subfigure}
  \hfill
  \begin{subfigure}[b]{.53\linewidth}
    \centering
    \includegraphics[width=\linewidth]{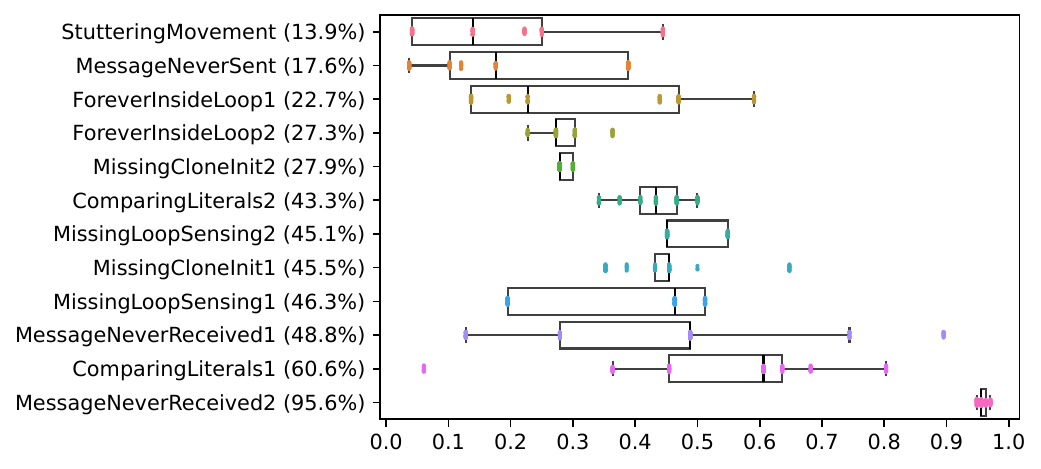}
    \caption{Per project, using \blockSus $\times$ \cumuChk}
    \label{fig:exams-dstar2-per-project}
  \end{subfigure}
  \vspace{-2em}
  \caption{Distribution of EXAM scores for DStar2 (median values in parentheses)}
\end{figure}

\Cref{fig:exams-dstar2-per-combination} shows the distribution of EXAM scores for different configurations using DStar2. The best median score of
\bestMedExamScore is achieved with \blockSus suspects and \cumuChk checking, meaning roughly one fourth of a program
must be examined before its fault is localized. It precisely deemed the faulty block most suspicious for 6 of the \nSimpleProjs
projects. Here, on average 13 other non-faulty blocks were equally suspicious. In contrast, the worst median
score of \worstMedExamScore is obtained by \targetSus and \testChk. The faulty block had the highest suspiciousness in
9~cases, but on average, \num{37.5} other non-faulty blocks had the same suspiciousness. Still, all techniques analyzed
yield a better EXAM score than a random selection of blocks, whose expected median is 0.5.

Overall, the best techniques use the \blockSus suspect level. We
conjecture this is because in our dataset the fault is caused by an
individual block, instead of an entire script or sprite. The
techniques with worse scores use coarser suspect levels, which makes
it difficult to isolate faulty blocks from correct parts of the
program. The best technique combines \blockSus with fine-grained
\cumuChk checking. We presume that unlike \assChk and \testChk, it
might be able to rectify the comparatively small number of tests in
\whisker test suites, allowing the SFL metric to provide more detailed
feedback.

\begin{figure*}
  \centering
  \begin{subfigure}[b]{.25\linewidth}
    \centering
    \begin{scratch}[scale=0.4]
      \blockinit{when \greenflag clicked}
      \blockrepeat{forever}{
        \blockif{if \boolsensing{key \ovalsensing*{left arrow} pressed?}}{
          \blockmove{change x by \ovalnum{-10}}
        }
      }
    \end{scratch}
    \caption{Fixed variant}
    \label{fig:stutter-movement-fixed}
  \end{subfigure}
  \hfill
  \begin{subfigure}[b]{.2\linewidth}
    \centering
    \begin{scratch}[scale=0.4]
      \blockinit{when \selectmenu{left arrow} pressed}
      \blockmove{change x by \ovalnum{-10}}
    \end{scratch}
    \caption{Buggy variant}
    \label{fig:stutter-movement-buggy}
  \end{subfigure}
  \hfill
  \begin{subfigure}[b]{.4\linewidth}
    \centering

    \begin{lstlisting}[gobble=6,basicstyle=\ttfamily\tiny]
      async function testCatMovement(p) {
        const cat = p.getSprite("Cat")
        const catX = cat.x;
        const input = {device: 'keyboard',
          key: 'Left', duration: 300};
        p.addInput(0, input);
        await p.runForTime(300);
        p.assert.less(cat.x, catX);
      };
    \end{lstlisting}
    \caption{\whisker test case exposing the bug}
    \label{lst:stutter-movement-test}
  \end{subfigure}
  \vspace{-0.5em}
  \caption{Excerpt from the \bestMedExamScoreProj project}
\end{figure*}

Finally, \cref{fig:exams-dstar2-per-project} considers the EXAM scores of the best combination \blockSus $\times$
\cumuChk per project in the dataset, revealing large differences. Fault localization worked best for
\bestMedExamScoreProj with a median EXAM score of \bestMedExamScorePerProj. The project implements a game, where a cat
is skiing down a mountain. Without carefully timed inputs from the player, the cat glides towards the right edge of the
screen. The task is to steer the cat in the opposite direction, avoiding as many obstacles as possible. The correctly
functioning implementation uses a forever-loop to monitor these inputs as shown in \cref{fig:stutter-movement-fixed},
making sure the cat moves left while the user is pressing and holding the arrow key. The buggy version uses the code in
\cref{fig:stutter-movement-buggy}. In particular, the event handler
{\setscratch{scale=0.4}
  \begin{scratch}\blockinit{when \selectmenu{left arrow} pressed}
\end{scratch}} only reacts to button taps, resulting in
stuttering movement, and should have been replaced with
{\setscratch{scale=0.4}
  \begin{scratch}\blockinit{when \greenflag clicked}
\end{scratch}}. This change also necessitates adding a forever-loop
and if-condition as in \cref{fig:stutter-movement-fixed} (fault of omission for
  {\setscratch{scale=0.4}
    \begin{scratch}\blockmove{change x by \ovalnum{-10}}
\end{scratch}} in). The test in \cref{lst:stutter-movement-test}
detects these problems, and FL correctly assigns the highest suspiciousness to both blocks in
\cref{fig:stutter-movement-buggy}.

On the other hand, \worstMedExamScoreProj had the worst EXAM score of \worstMedExamScorePerProj. Here, the player has
to navigate a gardener through a field of cacti, without touching any of them. Otherwise, the game is over, in which
case all remaining cacti glide towards the gardener. This is implemented by broadcasting a message
{\setscratch{scale=0.4}
  \begin{scratch}\blockevent{broadcast \ovalcontrol*{cactus touched}}
\end{scratch}} from the gardener to the cacti. In the
buggy version, these listen to the wrong message
{\setscratch{scale=0.4}
  \begin{scratch}\blockinit{when I receive \selectmenu{game over}}
\end{scratch}}, and thus the script implementing the
gliding is never executed. The analyzed SFL metrics assume only live code can cause a bug, while dead code never
can. This is why the block containing the bug is assigned the lowest suspiciousness. It is worth mentioning
the test suite for \worstMedExamScoreProj has similar block coverage as the other projects in \cref{tab:simple}. Yet,
all FL techniques perform poorly on it. This is evidence that bug finding ability is not just
impacted by suspect levels, checking levels, and coverage, but additional characteristics of Scratch programs not considered here.

\summary{RQ1}{The combination of DStar2, \blockSus suspects, and \cumuChk checking reveals faults best, but the choice of SFL metric
  is largely irrelevant. More fine grained suspect levels and checking levels tend to improve fault localization.
Code coverage is likely not a well-suited metric for how well tests exercise complex user-behavior.}

\subsection{\RQTWO}\label{subsec:results-rq2}

\begin{table}
  \scriptsize
  \caption{Number of fixes (full/partial) achieved on the \simple dataset}
  \vspace{-1em}
  \begin{tabular}{l*{6}{r@{\,/\,}r}}
\toprule
& \multicolumn{4}{c}{GA} & \multicolumn{4}{c}{RS} & \multicolumn{4}{c}{(1+1)EA} \\
\cmidrule(lr){2-5} \cmidrule(lr){6-9} \cmidrule(lr){10-13}
& \multicolumn{2}{c}{\fsSol} & \multicolumn{2}{c}{\fsInit} & \multicolumn{2}{c}{\fsSol} & \multicolumn{2}{c}{\fsInit} & \multicolumn{2}{c}{\fsSol} & \multicolumn{2}{c}{\fsInit} \\
\midrule
   ComparingLiterals1 &        15 &        15 &         15 &         15 &        15 &        15 &         15 &         15 &              8 &              8 &               8 &               8 \\
   ComparingLiterals2 &        14 &        15 &          0 &         15 &        15 &        15 &          0 &         15 &             12 &             15 &               0 &              15 \\
   ForeverInsideLoop1 &        15 &        15 &         15 &         15 &        15 &        15 &         15 &         15 &             15 &             15 &              15 &              15 \\
   ForeverInsideLoop2 &        15 &        15 &         15 &         15 &        15 &        15 &         15 &         15 &              5 &             14 &               3 &              14 \\
MessageNeverReceived1 &        11 &        15 &         11 &         15 &         6 &        15 &          8 &         15 &             11 &             15 &              11 &              14 \\
MessageNeverReceived2 &        15 &        15 &         15 &         15 &        15 &        15 &         15 &         15 &             14 &             14 &              14 &              14 \\
     MessageNeverSent &        13 &        13 &         12 &         12 &         0 &         0 &          0 &          0 &              9 &              9 &               6 &               6 \\
    MissingCloneInit1 &        11 &        15 &         11 &         15 &         0 &        15 &          0 &         15 &              2 &             15 &               3 &              15 \\
    MissingCloneInit2 &        15 &        15 &         15 &         15 &        15 &        15 &         15 &         15 &             13 &             13 &              15 &              15 \\
  MissingLoopSensing1 &        15 &        15 &         15 &         15 &        15 &        15 &         15 &         15 &             15 &             15 &              14 &              14 \\
  MissingLoopSensing2 &        15 &        15 &         15 &         15 &        15 &        15 &         15 &         15 &             12 &             12 &              10 &              10 \\
   StutteringMovement &        15 &        15 &         15 &         15 &        15 &        15 &         15 &         15 &             15 &             15 &              15 &              15 \\ \midrule
   Full Fix Rate &  \multicolumn{2}{r}{\num[round-mode=figures,round-precision=3]{0.938889}} &   \multicolumn{2}{r}{\num[round-mode=figures,round-precision=3]{0.855556}} &  \multicolumn{2}{r}{\num[round-mode=figures,round-precision=3]{0.783333}} &   \multicolumn{2}{r}{\num[round-mode=figures,round-precision=3]{0.711111}} &       \multicolumn{2}{r}{\num[round-mode=figures,round-precision=3]{0.727778}} &        \multicolumn{2}{r}{\num[round-mode=figures,round-precision=3]{0.633333}} \\
   Partial Fix Rate &  \multicolumn{2}{r}{\num[round-mode=figures,round-precision=3]{0.988889}} &   \multicolumn{2}{r}{\num[round-mode=figures,round-precision=3]{0.983333}} &  \multicolumn{2}{r}{\num[round-mode=figures,round-precision=3]{0.916667}} &   \multicolumn{2}{r}{\num[round-mode=figures,round-precision=3]{0.916667}} &       \multicolumn{2}{r}{\num[round-mode=figures,round-precision=3]{0.888889}} &        \multicolumn{2}{r}{\num[round-mode=figures,round-precision=3]{0.861111}} \\
\bottomrule
\end{tabular}

  \label{tab:fixes-rq2}
\end{table}

The experiments on fixing the faults in the \simple dataset used
DStar2 × \blockSus × \cumuChk for fault localization, as it emerged as
the best technique in \cref{subsec:results-rq1}. \Cref{tab:fixes-rq2}
summarizes the results: For each combination of repair subject,
algorithm, and fix source, it lists the number of full and partial
fixes achieved. The last two rows contain the fix rates computed
across all projects.  A recurring pattern emerges: The GA achieves the
highest fix rates, followed by RS, and (1+1)EA. Moreover, using the
\fsSol fix source yields more fixes than \fsInit. These observations
hold for both full and partial fixes.  For instance, GA × \fsSol
achieved the highest full and partial fix rates of
\bestFullFixRateSimple and \bestPartFixRateSimple, while (1+1)EA ×
\fsInit performed worst in terms of full (\worstFullFixRateSimple) and
partial fix rates (\worstPartFixRateSimple).

Interestingly, if \fsInit was used, no algorithm was able to fully fix ComparingLiterals2. It implements the well-known
game Pong. The player's score is kept in a variable \ovalvariable{my score}. Over time, the score increases, and
once a threshold of \enquote{1} is reached, the difficulty level increases. In the fixed version, this check is
implemented as \booloperator{\ovalvariable{my score} > \ovalnum{1}}, but the buggy version uses
\booloperator{\ovalnum{my score} > \ovalnum{1}}, comparing the literal \enquote{my score} against \enquote{1}, which is
always true. To fix this, \ovalvariable{my score} must appear elsewhere in the fix source, such that it can be copied
and inserted in place of the literal---but with \fsInit, this is not the case. This highlights the importance of blocks
available in the fix source.

\begin{figure}[tb]
  \centering
  \includegraphics[width=.85\linewidth]{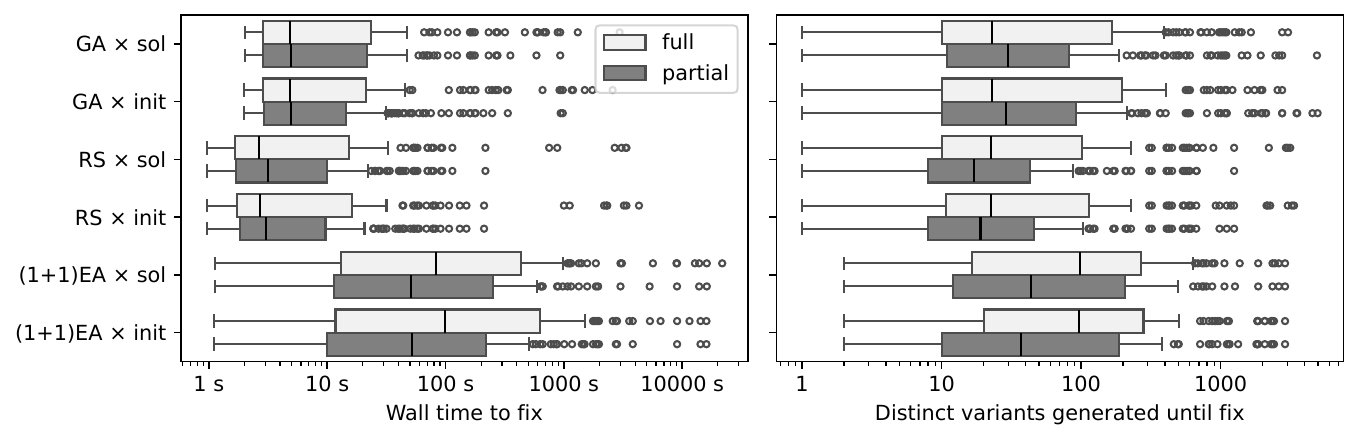}
  \vspace{-1em}
  \caption{Algorithm runtime and number of distinct variants generated until the first fix is found in the \simple dataset, considering only successful runs}
  \label{fig:time-variants-rq2}
\end{figure}

RS never produced a partial fix for MessageNeverSent, while the other algorithms were more successful with at least 6
full fixes. This project implements an animated story, and control flow is linear. Rather than having a single long
script, it contains several shorter ones, each starting with a
{\setscratch{scale=0.4}
  \begin{scratch}\blockinit{when I receive \selectmenu{message}}
\end{scratch}} hat. At the end of a script, another
one is triggered by broadcasting a message {\setscratch{scale=0.4}
  \begin{scratch}\blockevent{broadcast \ovalcontrol*{message}}
\end{scratch}},
similar to functions with tail calls in \enquote{traditional} languages. The buggy version is missing a tail call. This
can be fixed by copying the missing {\setscratch{scale=0.4}
  \begin{scratch}\blockevent{broadcast \ovalcontrol*{message}}
\end{scratch}} from
\fsSol, and inserting it at the right place. Because RS can only mutate the initial variant, a mutation has to get
this right immediately, the chance of which is very small.
Because \fsInit does not contain the needed block, it is impossible for RS to synthesize a fix. Yet, the GA and (1+1)EA
were able to work around this in some cases by incrementally copying blocks from one script to the other, effectively
concatenating them, the result of which is similar to inlining a tail call.

Despite the good fault localization on MessageNeverSent (see \cref{subsec:results-rq1}), RS failed to
insert the needed block at the right position. On the other hand, it produced a full fix for MessageNeverReceived2 in
all \nReps repetitions, despite bad EXAM scores. As explained in the previous section, the project already contains
the correct block,
and switching to a different message fixes the bug.

The GA excels at finding full fixes for MessageNeverSent and MissingCloneInit1, where the other algorithms tend to
struggle. Here, it is likely a full fix requires bigger changes than a single mutation, e.g., via crossover or by
successively mutating the project over several generations. While RS can only ever apply a single mutation to the
initial variant, the (1+1)EA has to keep evolving the current best variant. This strategy may lead to the required fix
eventually, but may cause other regressions along the way. In contrast, the GA might be able to better recover from
\enquote{bad} variants, thanks to it evolving an entire population of chromosomes.

\Cref{fig:time-variants-rq2} shows the distribution how long it took
each algorithm to find a fix, measured in wall time. Note that the
plot only considers successful runs, and excludes runs that did not
produce a fix. In general, if RS finds a fix then it tends to find it
fastest, followed by the GA. The runtimes for (1+1)EA are one order of
magnitude longer. This is because the other two algorithms are
parallelized and evaluate \nParallelEvals variants at a time, while
(1+1)EA evaluates just 1 at a time. To allow for a better comparison,
the figure also shows the number of distinct (syntactically different)
variants created until the first fix was found. Here, we observe the
same trend as with runtime: RS and the GA are most effective. While
the gap to (1+1)EA has shrunk, it still creates noticeably more
variants.

\summary{RQ2}{Random Search is quick and efficient at finding simple fixes. Evolutionary search takes longer, but is necessary for
more complex repairs.}

\subsection{\RQTHREE}\label{subsec:results-rq3}

\begin{figure}
  \centering
  \includegraphics[width=.85\linewidth]{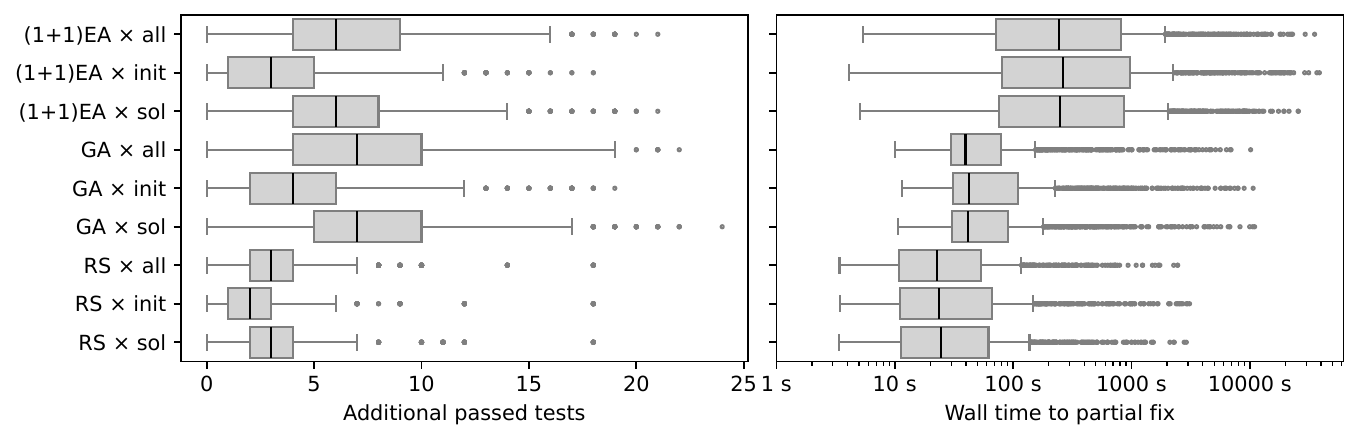}
  \vspace{-1em}
  \caption{Number of additional passed tests and algorithm runtime until first partial fix is found in the \complex dataset, and time until a partial fix was found in successful runs}
  \label{fig:additional-passed-time-rq3}
\end{figure}

All algorithms succeeded in providing at least a partial fix (i.e., at
least one more test passed than in the original program) for each of
the student solutions at least once, resulting in partial fix rates
close to \SI{100}{\percent}. The highest rate (0.977) was achieved by
GA × \fsSol, and the lowest (0.845) by RS × \fsInit.  Again, the GA
and fix sources containing the needed blocks lend themselves to fixing
bugs. However, even with \fsInit many partial fixes can be
found. Since a partial fix is already sufficient in order to provide a
hint to learners, this is an encouraging result.
However, \cref{fig:additional-passed-time-rq3} shows that the achieved
improvements in terms of the number of additional passed tests in the
resulting programs is again substantially larger for both search
algorithms compared to RS, with the GA producing the largest
improvements overall. Consequently, the solutions produced by the GA
are the most correct among the algorithms, which implies that
\emph{better} hints are likely produced by these algorithms.

On the other hand, only 11 of the 82 projects were fully repaired at
least once. The GA was most successful overall (fix rate of 0.0479),
followed by (1+1)EA. While RS still managed to find fixes for the
\simple dataset, it did not produce any full fixes on the \complex
dataset, because unlike the \simple dataset
it contains arbitrarily broken programs. To fully fix a project,
it is likely bigger changes are required than RS is able to
produce with individual mutations. For the other two algorithms,
pairing them with \fsSol leads to the highest repair success, while
the least fixes were produced with \fsInit. This is likely because
\fsSol contains the missing blocks needed for a fix.  The \fsAll fix
source performs slightly worse than \fsSol.  Besides the fixed
variant, it also contains other buggy variants, which might introduce
additional unwanted broken code.
Overall, this suggests the plastic surgery hypothesis indeed does
not hold on learners' \Scratch programs, but (1) even without
it, a genetic search can feasibly produce
fixes, and (2) the plastic surgery hypothesis can be recovered using
other data sources, such as model or student solutions, increasing the
efficiency of the search. Consequently, it is important that educators
provide model solutions to improve automated hints.

Besides the ability to generate fixes, we also consider how timely
feedback can be provided.  \Cref{fig:additional-passed-time-rq3} plots
the time each algorithm takes to find the first partial fix. The
figure suggests RS finds a fix fastest again, followed by the GA. As
with the \simple dataset, (1+1)EA is one order of magnitude slower,
due to not being parallelized. While fixes seem to be found fastest
with the \fsAll fix source, choosing another fix source increases the
time only marginally. While this means there is a slight
trade-off between the correctness of solutions produced and the time
until the first improvement is found, we argue that considering the
runtimes a hint generation system would more likely need to integrate
any of the repair algorithms in a controlled manner (e.g.,
providing feedback after submission, or at fixed intervals). There is,
however, ample potential for improvements in future research to enable
interactive use in the IDE.

One immediate application that we plan to study is to use repair to create a larger
pool of model solutions (by repairing solution attempts), to improve existing hint
systems. In this scenario, the performance of the repair is irrelevant, since it can be
done offline.
Furthermore, we anticipate that
\begin{enumerate*}
\item the same student may ask for hints on same/similar solutions many times, and
\item many learners may produce similarly broken solutions.
\end{enumerate*}
Evolutionary search provides potential for speed up by gradually improving fix sources
or by using population seeding.
Our approach represents a baseline search-based repair, and establishes that repair is
possible in this domain. Given this foundation, future work can investigate
optimizations such as
\begin{enumerate*}
\item surrogate models to speed up fitness calculations,
\item integrating machine learning, e.g. cloze-style repair, or
\item integrating LLMs.
\end{enumerate*}

\summary{RQ3}{Partial fixes can almost always be found, and random
  search is quickest at doing so. Evolutionary search, however,
provides substantially larger improvements of correctness.}


\section{Related Work}\label{sec:related-work}

Automated program repair (APR) is a thriving area of software
engineering research, popularized initially by GenProg~\cite{genprog}
and other early tools, and with substantial progress over
time~\cite{gazzola2019}. APR approaches are usually classified into
heuristic repair, constraint-based repair, and learning-aided repair
approaches~\cite{Goues2019}. Recently, the focus of research has been
specifically on learning-based repair~\cite{zhang2023}. Our tool uses
a heuristic repair approach with a genetic algorithm similar to
GenProg, rather than alternative heuristic approaches based on
templates (e.g.,~\cite{Kim2013,Liu2019}), since fix-commits are not as
readily available for \Scratch for mining such patterns
(e.g.,~\cite{Koyuncu2020}) in contrast to traditional programming
languages. Similarly, learning-based repair approaches are currently
still inhibited by a lack of adequate models for block-based \Scratch
programs~\cite{Griebl2023}, but are certainly an avenue for further
research on improving the search-based repair~\cite{Wang2018b}.

Different application scenarios are considered in the APR literature,
and repairing student programs is a common one: For example, a number
of established APR benchmarks consist of collections of student
programs~\cite{tan2017,Goues2015,Nakamura2022}.
However, empirical results have indicated that out-of-the-box APR
tools do not perform well on student submissions~\cite{Yi2017}. This
has led to the development of dedicated APR approaches for the
scenario of repairing student programs, such as
Clara~\cite{Gulwani2018}, SarfGen~\cite{Wang2018},
ErrorCLR~\cite{Han2023}, Refactory~\cite{Hu2019},
AssignmentMender~\cite{Li2023}, TipsC~\cite{Sharma2018},
sk{\_}p~\cite{Pu2016}, AutoGrader~\cite{Singh2013}, and
others~\cite{Marin2017,Wang2018b}.
A number of repair approaches also specifically target syntax errors
rather than functional errors
(e.g.,~\cite{Parihar2017,Ahmed2018,BhatiaS16,Gupta2017,Gupta2019}).
But none of these approaches considers the specific context of
block-based programs as in \Scratch, which
creates unique challenges as described in this paper.
In the context of block-based programming, next-step hints are usually
based on data-driven techniques rather than automated program
repair. For example, Catnip~\cite{Fein2022} for \Scratch and
iSnap~\cite{Price2017} for Snap! match syntactically similar
alternative solutions, as often available in classroom or MOOC
scenarios, and synthesize edit diffs~\cite{zimmerman2015}.

Several of the specific challenges and improvements implemented in our
tool and evaluated in our study match problems known from other
domains, often exacerbated by the specific nature of \Scratch
programs.
For example, search-based repair tends to generally suffer from
insufficient guidance provided by common fitness function, leading to
attempts for improvement~\cite{Fast2010,Souza2018,Bian2021}; our
proposed fitness function generalizes beyond \Scratch in
principle.
The question of the search algorithm to use is another central
question in APR research, and our findings confirm prior studies:
Random search is quick and effective, but applying a genetic algorithm
leads to better and more repairs~\cite{Qi2014,Assiri2015}.
The importance and challenges of fault localization for APR have been
demonstrated~\cite{SoremekunKBP23}, and it is known that student
programs are particularly challenging for fault localization, leading
to specifically tailored suspiciousness metrics~\cite{Li2021}. The
assertion-based fault localization introduced in this paper is not
specific to \Scratch and \whisker tests and in principle generalizes
to all automated tests.
In the future, we plan to investigate the use of automatically generated
tests. While prior research shows a degradation of the repair
effectiveness~\cite{Motwani2022}, automatically generated tests so far
mainly served to address~\cite{Yang2017,Xin2017} the
problem of patch overfitting~\cite{smith2015}.
%
Finally, a related challenge in APR is the negative impact of flaky
tests~\cite{Qin2021}, which we addressed by using a deterministic
modified virtual machine for executing \whisker
tests~\cite{scratch-automated-test-generation}.

A hypothesis underlying this paper is that the fixes generated for
learners' programs are a requirement for providing useful feedback. The question of how to
actually synthesize adequate feedback for learners is a complementary
question investigated in the computer science education
community~\cite{Hao2022,Obermueller2023,Marwan2019,Hao2019}; however,
a fix is always a useful starting point for generating this feedback.
Additionally, the ability to evolve \Scratch programs has further
applications beyond feedback generation; for example, the search can
be applied to the orthogonal problem of generating new programming
assignments~\cite{Ahmed2020}, or synthesizing new programs with faults
similar to those made by students~\cite{Singla2022}.


\section{Conclusions}\label{sec:conclusions}

Automated program repair is an important ingredient for automated
feedback to programming learners. Learners are increasingly introduced
to programming using block-based programming languages, for which
automated program repair has not previously been explored. Even though
the programs may look simple and playful, they pose many challenges
for automated program repair, as they may be inherently incomplete,
contain many faults, are not supplied with thorough test suites that
can precisely locate bugs or guide the search for fixes, and do not
generally provide redundant code suitable to satisfy the plastic
surgery hypothesis. In this paper we faced these challenges and
introduced and evaluated the first APR approach for \Scratch,
demonstrating feasibility.

Although our approach successfully manages to overcome many of the
existing challenges and violated assumptions, our experiments reveal a
central issue that may be surprising, considering the small size of
the relevant programs: Long running tests cause substantial
computational costs for automated program repair. There are, however,
multiple ways forward to address this issue. For example, there are
many promising avenues for hybridizing our search-based approach with
alternatives: Data-driven hint generation
techniques~\cite{zimmerman2015} could help selecting fix sources and
code for copying; similarly, cloze-style learning-based approaches
could represent suitable alternative fix sources given language models
trained on \Scratch~\cite{Griebl2023}; large language models may be
helpful in cutting down the search time~\cite{XiaWZ23}; evaluation of
candidate repairs can be optimized by selecting subsets of test
suites~\cite{MehneYPSGK18}, or by only partially executing \whisker
tests. Finally, while we evaluated individual and independent repair
attempts, in practice the same student may request feedback for their
programs multiple times after only small modifications, there may be
many similar solution attempts simultaneously or across cohorts, and
partial repairs may be sufficient for providing feedback, such that a
full repair run may actually only sporadically be required.

Besides the question of improving the performance, future research
will also focus on the complementary problem of synthesizing actual
feedback from the generated repairs~\cite{Hao2022}, as well as
alternative applications of the search algorithm, such as for
producing model solutions, alternative solution paths for existing
programs, or entirely new programming tasks.

\section*{Data Availability}

We provide implementations, the raw data obtained during our experiments, and the
evaluation scripts to generate the plots in this paper at
\url{https://doi.org/10.6084/m9.figshare.27014959.v1}.

\section*{Acknowledgements}

This work is supported by BTHA project BTHA-JC-2024-41. We thank the reviewers for
their comments and suggestions, which helped to improve this manuscript.


\bibliographystyle{ACM-Reference-Format}
\bibliography{bibliography}

\end{document}